\documentclass[runningheads]{llncs}

\usepackage{graphicx}
\usepackage{color}
\usepackage{booktabs}
\usepackage{subcaption}
\newcommand\nwonote[1]{}
\newcommand\ignore[1]{}

\begin{document}
\title{Coordinated Information Campaigns on Social Media: A Multifaceted Framework for Detection and Analysis}
\titlerunning{Coordinated Info Ops: A Framework for Detection and Analysis}
%
\author{Kin Wai Ng\inst{1}\orcidID{0000-0001-9784-8427} \and
Adriana Iamnitchi\inst{2}\orcidID{0000-0002-2397-8963}}
\authorrunning{K. Ng et al.}
%
\institute{University of South Florida, Tampa FL 33620, USA\\ 
\email{kinwaing@usf.edu}
\and
Maastricht University, Maastricht, The Netherlands\\
\email{a.iamnitchi@maastrichtuniversity.nl}}
\maketitle              
\begin{abstract}
The prevalence of coordinated information campaigns in social media platforms has significant negative consequences across various domains, including social, political, and economic processes.
 This paper proposes a multifaceted framework for detecting and analyzing coordinated message promotion on social media.
 By simultaneously considering features related to content, time, and network dimensions, our framework can capture the diverse nature of coordinated activity and identify anomalous user accounts who likely engaged in suspicious behavior.
 Unlike existing solutions that rely on specific constraints, our approach is more flexible as it employs specialized components to extract the significant structures within a network and to detect the most unusual interactions.
 We demonstrate the effectiveness of our framework using two Twitter datasets, the Russian Internet Research Agency (IRA), and long-term discussions on Data Science topics. 
 The results demonstrate our framework's ability to isolate unusual activity from expected normal behavior and provide valuable insights for further qualitative investigation. 
\keywords{Coordinated Campaigns \and Information Operations \and Social Media}
\end{abstract}

\section{Introduction}
\label{sec:intro}

Social media platforms have been under scrutiny for allowing nefarious processes on their sites. 
Information  coordination campaigns are such processes that can inflict significant damage on the society. 
Implemented as disinformation campaigns~\cite{wilson2021cross,kin2021multi}, social activism~\cite{merlan22activism}, elections~\cite{lukito2020coordinating}, or digital currency manipulation~\cite{nghiem2021detecting}, they appear as organic, spontaneous conversations among unrelated user accounts who post different messages in support of the same agenda within a short time interval. 
Such accounts are not necessarily bot accounts, but very often verified and even influencer accounts\footnote{\url{https://www.vice.com/en/article/akewea/a-pr-firm-is-paying-tiktok-influencers-to-promote-liberal-causes-and-hype-democrats-middling-accomplishments}}, escaping thus bot detection tools.  
The messaging promoted is not always in bad faith, as it can be part of health promotion campaigns or legitimate political campaigns. 
However, recognizing such coordinated campaigns and distinguishing them from organic social media interactions and message sharing is important both for understanding the media landscape and for limiting manipulation. 

There are many challenges in addressing the problem of identifying coordinated information operations. 
First, it is a relatively rare event, thus it is difficult to learn enough about such practices and difficult to detect automatically (the ``needle in the haystack'' problem). 
Second, while temporal locality may be a requirement for the success of such operations, in practice it is unclear how to select a representative time window relevant for platforms with different posting frequencies. 
For example, what does time coordination look like when a message is introduced in an ongoing Reddit discussion thread vs. in a tweet?
Third, publicly available data on information campaigns is very limited, with little to no information about what constitutes actual patterns of inauthentic coordinated behavior.
Thus, distinguishing between organic, synchronized behavior and coordinated campaigns is a difficult task. 

Coordinated message promotion is characterized by locality in content and time, and similarity in the activities of the user accounts involved. 
Therefore, attempts to detect such campaigns focus on identifying unusual patterns related to time, content, and user activity.
For example, some solutions assumed predefined thresholds on time~\cite{giglietto2020detecting,pacheco2021uncovering,kin2021multi} or on content similarity~\cite{lynnette2022cross,lee2014campaign} to separate organic from potentially suspicious actions. 
Moreover, many of the previous solutions functioned as a pipeline in which the time, content and user actions (often modeled as network anomalies) were detected in a sequential order. 
However, predefined thresholds are somewhat artificial and easy to bypass. 

We propose a methodology that avoids fixed time or similarity thresholds and can generalize to different social media platforms.
Our solution starts by building a network of user accounts connected by weighted edges that quantify similar interest in content posted.
We then reduce this very dense network to its backbone, a procedure which maintains only the edges that represent the higher information similarity.
Information similarity, in our case, is measured as vocabulary overlap of sets of posted messages.
On the resulting backbone network we run a community detection algorithm to identify clusters of user accounts with stronger connections. 
In each cluster, we identify the pairs of accounts that deviate the most from the pair-wise activity of the other accounts in the cluster in terms of timing as measured by inter-arrival time between actions, content similarity as measured by the cosine similarity between text embeddings, and network similarity as measured by the cosine similarity of node embeddings.
It is this subset of anomalous user accounts that we believe should be studied via qualitative methods to reliably identify coordinated campaigns. 
Our solution, like~\cite{francois2021measuring}, looks simultaneously at all the three dimensions necessary for a coordinated information campaign: time synchronicity, content locality and coordinated user activity. 
However, we define a different network than their follower-followee network in an attempt to provide a platform-independent solution. 
Our solution could augment previous solutions tailored for particular platforms.

\section{Related Work}
\label{sec:related}

Previous studies have proposed different frameworks to identify coordination among actors by examining platform-specific features such as co-retweet or co-tweet networks~\cite{keller2020political}, or the retweet network~\cite{gupta2019malreg}.
Similar to our work, other studies have identified coordination via shared pieces of content, such as topics, narratives, hashtags, or URLs.
For example, Vargas et al.~\cite{vargas2020detection} utilized network statistics to predict coordinated activity on Twitter by defining coordination behavior in terms of co-shares.
Giglietto et al.~\cite{giglietto2020detecting} identified groups of coordinated accounts based on co-shared URLs.
Pacheco et al.~\cite{pacheco2021uncovering} proposed a generalized, qualitative approach for detecting coordinated behavior by exploiting behavioral traces (e.g., temporal patterns of activity) or common actions (e.g., sharing the same content or URLs) to identify groups of coordinated campaigns.
Weber et al.~\cite{weber2020s} proposed an approach to detect coordinated groups based on latent connection networks and focal structure analysis.
More recently, Magelinski et al.~\cite{magelinski2022synchronized} focused on constructing a multi-view network using common interactions to uncover synchronized actions within narrow time windows.
A limitation of these frameworks is that they rely on discrete time windows or predefined short time thresholds to detect coordinated instances, which can potentially result in missing instances of coordination in more intricate and adaptive campaigns. 

Only a few studies have focused on developing frameworks that specifically target the identification of coordinated campaigns by considering several dimensions such as network, time, and content. 
Kriel et al.~\cite{kriel2019reverse} studied the IRA dataset using network analysis to investigate the temporal evolution of network content pushed by Twitter bots and accounts related to online influence campaigns.
The framework proposed by Francois et al.~\cite{francois2021measuring} shares some similarities with our proposed methodology as it also focuses on identifying coordinated activities through the analysis of network, temporal and semantic dimensions.
However, their approach relies on constructing networks based on follower-followee interactions, which are specific to some platforms only, and can be challenging or very expensive to obtain. 

In this work, we proposed a framework that also investigates coordinated activity through the analysis of three different axes: content, timing and network structures.
Unlike previous work, our approach leverages techniques to extract the key structural components of the network and employs an unsupervised machine-learning model for effective anomaly detection. 
The objective of our framework is to isolate unusual activity from what is considered normal behavior in a particular context.
We assume that most of the social media activity under a specific context/discussion follows a normal pattern, while coordinated interactions will exhibit distinct characteristics that set them apart.
This observation typically holds true even in datasets directly associated with an information campaign.
For example, in the IRA dataset, accounts identified as part of the campaign were found to frequently share banal content, including sports or local news, along with trending hashtags to inject themselves into popular discussions and gain followers~\cite{kriel2019reverse}.

\section{Methodology}
\label{sec:method}

Our approach to identifying potentially coordinated information operations is based on the following intuition: in order for a message to reach a large number of people, it has to be repeated within a short interval of time by multiple apparently unrelated user accounts. 
Thus, a coordinated information operation requires content and time locality, where content locality means possibly distinct messages in support of a shared objective. 

Our solution is based on the following observations from forensic studies of information campaigns~\cite{bellutta2023investigating,wilson2021cross,francois2021measuring,pacheco2021uncovering,sharma2020identifying}. 
First, accounts involved in a coordinated campaign will exhibit persistent behavior: the same account will post repeatedly on the same topic to promote a message. 
Second, multiple accounts will participate in promoting the same narrative for a successful (and thus, worth identifying) campaign. 
Third, we assume that users who are engaged in a coordinated information operation are likely to have similar tasks to perform, which might translate into similar network structures or connection patterns.
Fourth, we implicitly assume that normal behavior is more common than coordinated behavior. 

We propose a methodology that consists of five stages, namely \textit{network construction}, \textit{backbone extraction}, \textit{community detection}, \textit{feature extraction}, and \textit{anomaly detection}. 
Briefly, we first construct a network of user accounts connected by posts with similar content. 
In order to provide a platform-independent solution, we ignore resharing activities (typical of Twitter and LinkedIn, but not typical of YouTube or Reddit, for example) and only consider the original posting activities (e.g., posts in Reddit, tweets in Twitter, etc.). 
From this potentially large and quite dense network, we extract its backbone to ignore the user connections that are less active or less similar in the content promoted. 
We then detect network clusters that, due to the network construction methodology, will map onto shared topics in the promoted content.
We extract features that capture content, time, and network similarity among user connections in each cluster.
Finally, we identify anomalies from ``typical'' user behavior based on the observed features.
We assume that the frequency of ``normal'' activities in a particular context will be higher than that of coordinated activities as even coordinated accounts will occasionally engage in ``normal'' behavior to avoid detection.
Each component is described below.


\subsection{Co-sharing Network Construction}
\label{sec:cosharing}

The objective of this component is to identify shared interests among users based on the topics/information they post. 
We construct networks among social media user accounts based on the co-occurrence of similar pieces of information. 
We can define similarity of information in different ways, from identical URLs or hashtags to content-based analysis revealing the same topics or narratives. 

The co-sharing network is defined as a bipartite graph, $B = (U, V, E)$, where the nodes $u \in U$ represent social media accounts, and nodes $v \in V$ represent pieces of information.
An edge $e \in E$ between $u$ and $v$ refers to the number of times a social media user shared a particular information entity. 
We project this bipartite graph onto the social media user nodes to obtain an undirected graph consisting of user co-occurrence connections.
The edge weight between two users refers to the minimum number of times that both users were observed sharing the same pieces of information. 
As an attempt to reduce false positive connections (i.e., user co-shares happening by chance), we filter out those edges with an edge weight of 1.
This includes user pairs who are only seen sharing a piece of information once over the entire period.
In this study, we focused on using hashtags as the piece of information to connect users for our analysis. 

\subsection{Network Backbone Extraction}

The goal of this component is to identify and extract the most relevant accounts and their connections from the original network, aiming to eliminate accounts that do not often contribute on shared topics. 
Projections of bipartite networks 
lead to very dense structures where many of the edges are possibly affected by infrequent ties between the different node  types in the original network, which in our context may be seen as noise (e.g., spontaneous reactions to particular news or real-world events).
To address this challenge, previous work has adopted global threshold approaches where edges with weights higher than some threshold are kept while all others are removed~\cite{pacheco2021uncovering,lee2014campaign}.
This approach is not ideal for networks with skewed weight distributions, as it is often the case for social media networks.
Global thresholding techniques do not consider the multi-scale nature of such networks and thus relevant structures and hierarchies are overlooked.
Instead, a better strategy is to focus on locality, where the salient core network structure is decided at the node level.

In this study, we apply the Noise-Corrected (NC) backbone strategy proposed by Coscia et al.~\cite{coscia2017network}. 
While several backbone approaches have been proposed~\cite{serrano2009extracting,grady2012robust}, the NC backbone method is considered a more robust approach for identifying important edges in a network.
This is because it can reduce the occurrence of spurious correlations by comparing edge weights at the level of node pairs rather than at the level of individual nodes.
Unlike other methods, the NC backbone is capable of more effectively preserving the underlying topological characteristics of the original network while filtering out noisy connections.

The NC backbone uses a null-model based on the assumption that edge weights in the network are drawn from a binomial distribution.
The Bayesian framework is used to estimate the expected value and variance of edge weights while considering the likelihood of pairs of nodes to send or receive edges.
An edge is kept in the backbone if and only if its observed weight is higher than $\delta \sqrt{V[L_{ij}]}$, where $V[L_{ij}]$ is the estimation of the expected variance for the observed edge weight $L_{ij}$ between node $i$ and $j$, and $\delta$ is a parameter for the tolerance of noise in a particular network.
We set this value to 2.32 which approximates p-values at a significance level of 0.01, as suggested by the authors.

\subsection{Community Detection}

The objective of this component is to enable the selection of groups of users that exhibit overlapping content-based interests.
Specifically, this component involves identifying the communities of users within the backbone of the co-share network. 
We employ the Louvain algorithm for community detection, which is frequently used in prior related research~\cite{nizzoli2021coordinated,morstatter2018alt,nasim2018real}.
The Louvain algorithm works by optimizing a modularity score.
It measures the strength of the communities detected by comparing the density of connections among nodes in a given network with that in a random network. 
We accounted for edge weights in the Louvain algorithm, which enables the identification of communities based on the strength of links between users instead of just their presence.

\subsection{Edge Feature Extraction}
\label{sec:features}

We focus on extracting edge features related to content, time, and network dimensions.
These dimensions have been highlighted in prior research as critical factors for detecting coordination phenomena in social media platforms~\cite{francois2021measuring,bellutta2023investigating}.
Content features capture the similarity of content being shared between users, while temporal features capture the timing of their interactions.
Network features capture the structural role of users within the network. 
Our assumption is that users who engaged in coordinated operations are likely to perform similar tasks, which might translate into similar connections patterns or network structures. 
It is important to note that the edge features we extract in this study are not intended to be comprehensive as there may be additional features that could be relevant.
However, the edge features we consider are particularly relevant for capturing coordinated activity as they provide a general view of the interactions between users in the network.

\textbf{Edge weight} 
measures the propensity of two users to share similar pieces of information. Specifically, the edge weight is computed by considering the total frequency of co-occurrence of shared elements such as hashtags, URLs, or keywords being shared by two users. It is computed as follows:

\begin{equation}
    W_{ij} = \sum_{n=1}^{N} \min [\sigma(i, n), \sigma(j, n)]
\end{equation}

where $\sigma(i, n)$ denotes the number of messages posted by user $i$ that contain a given element $n$.
This feature captures the degree to which users are sharing/promoting similar pieces of information.

\textbf{Content similarity}
measures the degree of similarity in the content posted by two users. 
In this study, we focus on measuring content similarity between users' posts with the same hashtags.
We use the cosine similarity measure to compute the similarity between the embeddings of the text found in posts by two users sharing the same hashtag.
To obtain one value of content similarity between two users, we take the average of the cosine similarity values computed over all the hashtags that both users have in common.
The formula is as follows:

\begin{equation}
    C_{ij} = \frac{\sum_{n=1}^{N} cosineSimilarity[\sigma(i, n), \sigma(j, n) ]}{N}
\end{equation}

where $\sigma(i, n)$ denotes the average embedding vector of tweets posted by user $i$ under hashtag $n$.
The embeddings are extracted using a pre-trained sentence transformer model~\cite{reimers2019sentencebert}, specifically the \textit{paraphrase-multilingual-mpnet-base-v2} model\footnote{\url{huggingface.co/sentence-transformers/paraphrase-multilingual-mpnet-base-v2}}, which maps text in multiple languages to a 768 dimensional vector representation.
We chose this model because it was fine-tuned for sentence similarity tasks; thus, its vector representations are better for capturing semantic textual similarity.
While there are language models trained on Twitter data, they are primarily designed for tasks other than sentence similarity, and few of them incorporate multilingual data. 

\textbf{Temporal signature}
measures the timing between the posts of two users under the same hashtag. 
Specifically, we compute the shortest $\delta$ interarrival times, where $\delta$ is the number of co-shares between two users for a given hashtag.
The resulting distribution of interarrival times is summarized by taking the median interarrival time as the final value to represent timing for the particular user pair.
Unlike the mean, which is very sensitive to outliers, the median offers a more robust estimate of central tendency.

\textbf{Node similarity}
captures the similarity between two users in terms of their respective network position or structural role.
This feature is measured through computing the cosine similarity between the node2vec~\cite{grover2016node2vec} embeddings of two users.
Node2vec is a graph representation learning algorithm that maps nodes in a network to a low-dimensional embedding vector that captures their structural properties.
The choice of node2vec is motivated by the assumption that users who are engaged in coordinated activity are likely to have similar network structures or connection patterns.
Network similarity can provide a good proxy to detect coordinated activity between users, even when there is no explicit content similarity between their posts.

\subsection{Anomaly Detection}

The objective of this component is to isolate organic from inorganic behavior, with the assumption that organic behavior is more common and inorganic behavior (reflected by coordinated operations) will stand out.  
In this study, we employ Isolation Forests~\cite{liu2008isolation} for anomaly detection in the context of coordinated behavior in social media.
The algorithm consists of multiple binary decision trees trained with different subsamples drawn from the original data.
During training, each decision tree decomposes the data space into two subtrees using the arbitrary values of randomly selected features.
Isolation forests measure the degree of anomaly of a particular data instance by computing its average path length from the root of the tree.
The idea is that anomalous samples should require less effort to separate from the rest of the samples, which results in shorter path lengths across the trees.
Particularly, we trained an isolation forest for each identified cluster from the community detection stage by considering their respective edge characteristics as input features.

The anomalous data instances identified by the isolation forests consist of edges with significantly different characteristics from the normal distribution of edges in each cluster. 
This approach allows an understanding of what is considered normal and abnormal in a given distribution, thus providing important insights into coordinated behavior.
The results from the isolation forests can be combined with explainable methods such as SHAP~\cite{lundberg2017unified} to understand what features contribute the most to these anomalies.
The identification of anomalies enables a more focused investigation of behaviors with high indication of coordination.

\section{Datasets}
\label{sec:analysis}

Our analysis focuses on two datasets collected from Twitter.
The first is related to social media manipulation operations, and the second covers general topics of discussion related to technology trends.
We selected these datasets to cover a span of known coordinated information operations to likely only organic discussions.
For each dataset, we report basic summary statistics and information related to data collection and pre-processing.
We focus only on original tweets containing at least one hashtag.
This is because hashtags often indicate the general topic/theme of a tweet, and are often employed to boost a particular issue or narrative, especially in coordinated campaigns.

\textbf{Russian Internet Research Agency (IRA)}
This dataset consists of a subset of accounts that Twitter has identified as being linked to the Russian Internet Research Agency. 
The corpus of tweets and corresponding metadata, posted between 2009 and 2018 by these accounts, has been publicly released as part of the Twitter Election Integrity dataset\footnote{\url{https://transparency.twitter.com/en/reports/moderation-research.html}}.
We narrow our focus on the period between July 7th 2014 to November 31st 2016 since it contains several real-world events, such as the 2016 U.S. presidential election or the downing of the Malaysian airplane flight in Ukraine, which have been shown to be subject of significant intervention from the IRA in online discussions~\cite{golovchenko2018state,lukito2020coordinating}.
In total, the dataset contains 1,577,082 tweets on 18,826 unique hashtags from 3,594 users. 
For each hashtag, the number of tweets ranged from 2 to 236,322 with an average of 111 tweets per hashtag.
The number of unique users per hashtag ranged from 2 to 1,143 with an average of 11.5 users per hashtag.

\textbf{Data Science Tweets (DS)}
This dataset is a collection of tweets related to the trends and advancements in the field of data science over the past decade. 
It includes tweets that mention data science, data visualization, or data analysis.
The dataset is publicly available on Kaggle\footnote{https://www.kaggle.com/datasets/ruchi798/data-science-tweets}.
We focus on tweets over the period of January 1st 2016 to June 19th 2021.
In total, there are 142,282 tweets from 5,730 users and with 7,521 unique hashtags.
To allow for a better comparison in coordinated behavior within different contexts, we subsample the user accounts in the DS dataset to match the number of users in the IRA.
We ensure that the distribution of activities of the subsampled users approximates the distribution of user activity rates in the IRA dataset.
The resulting sampled user accounts are more representative of the activity levels of users in the IRA, and thus more comparable across the two datasets.
Overall, the sampled dataset contains 136,429 tweets on 7,192 unique hashtags from 3,594 users.

Each dataset includes the following fields: a unique identifier for the author of the tweet, a unique identifier for the tweet, the timestamp indicating when the tweet was posted, the text of the tweet, and a list of hashtags used in the tweet.
For each dataset, we removed user accounts with only one tweet over the entire period of the dataset, as they are unlikely to provide useful information for coordinated activity. 
We removed duplicated tweets (i.e., multiple occurrences of the same tweet ID, but not tweets with identical content).
We cleaned the tweets by removing mentions, URLs, and hashtags from the text to keep only the natural language content that likely reflects the user's opinion.
We use the \texttt{langdetect}\footnote{https://pypi.org/project/langdetect/} Python library to detect the language of the tweets in each dataset.
The proportion of non-English tweets within each dataset was 39.8\% in the case of the IRA, and 2.9\% in DS. 
We use a pre-trained many-to-many multilingual model, namely \textit{facebook/m2m100\_418M}\footnote{https://huggingface.co/facebook/m2m100\_418M}, to translate non-English tweets to English.
The translated text is only used for the qualitative analysis of non-English tweets.
For extracting tweet embeddings, we make use of the original text since in most cases the true meaning of a piece of text is lost in translation.

\section{Results}
In this section, we present the results of our framework as applied to the two datasets in this study. 
We investigate the impact of the backbone extraction component on reducing the size of the network and highlighting relevant structures.
We explore how features related to time, content, and network dimensions contribute in identifying anomalous instances.
Finally, we conduct a qualitative analysis of the anomalous clusters identified by our method, aiming to characterize each cluster based on their shared content and edge features.

\subsection{Extracting the Backbone of Co-sharing Networks}

As mentioned in Section~\ref{sec:cosharing}, the co-sharing networks were constructed by connecting user accounts based on shares of the same hashtags.
The total number of co-sharing interactions was 1,897,678 for the IRA and 3,199,919 in Data Science.
To reduce spurious co-sharing interactions, all edges with weight of 1 were removed from the original networks as they do not necessarily indicate coordinated behavior. 
The proportion of edges removed from the co-sharing networks was  21.5\% for the IRA, and 33.4\% for the Data Science dataset.
We applied the NC backbone strategy to each of these networks.
Table~\ref{tab:backbone_data} presents a comparison between the original networks and their backbones across several graph measures.
The backbone strategy reduces significantly the size of edges in the original co-sharing networks, while still preserving important network structures. 
Particularly, the proportion of edges removed from the original networks was 59\% in the IRA and 88\% in Data Science. 
We observed that the original networks exhibit higher density and centralization scores than their respective backbones. 
The centralization scores of the original networks are 3 and 4.9 times higher than their backbone in IRA and Data Science, respectively. 
This indicates that the original networks tend to be centralized around a small set of nodes with a high concentration of shared hashtags.
The backbone strategy reduces the centralization score by removing some particular connections to hubs, which are considered less important as hubs have a tendency to connect to a large number of nodes in the network.
This is also observed in the mean node degree of the backbone networks, which on average decreases by a factor of 5 compared to the original networks.

The backbone networks are capable of highlighting the underlying structures of the original graphs as seen by the increase in modularity scores, and preserving the most relevant interactions as evidenced by the increase in the average edge weight.
We ran Louvain community detection on the backbone networks to detect strongly connected clusters of users.
The algorithm identified 6 clusters in the IRA ranging from 37 to 937 users, and 5 clusters in Data Science from 135 to 1,661 users.

\begin{table*}
    \centering
    \caption{Network summary statistics for the original co-sharing networks and their corresponding backbone. Edges with a weight of 1 are omitted.}
    \begin{tabular}{lrrrr}
    \toprule
    & \multicolumn{2}{c}{\textbf{IRA}} &\multicolumn{2}{c}{\textbf{Data Science}}\\
    \midrule
    & \textbf{Original} & \textbf{Backbone} & \textbf{Original} & \textbf{Backbone} \\
    \textbf{Nodes (\#)}       &3,575 &3,575 &3,421          &3,421 \\
    \textbf{Edges (\#)}       &1,489k &604k  &2,131k         &249k  \\
    \textbf{Density}          &0.23 &0.09 &0.36           &0.04  \\
    \textbf{Centralization}   &0.46 &0.15 &0.54           &0.11  \\
    \textbf{Modularity}       &0.58 &0.65 &0.16           &0.43  \\
    \textbf{Mean Edge Weight} &32.7$\pm$158 &62.7$\pm$239 &6.7$\pm$37   &12.1$\pm$106 \\
    \textbf{Mean Node Degree} &833$\pm$511 &338$\pm$162 &1246$\pm$888 &146$\pm$95 \\
     \bottomrule 
    \end{tabular}
    \label{tab:backbone_data}
\end{table*}

\subsection{Anomaly Detection Using Isolation Forest}

\begin{figure*}[htbp]
    \centering
    \begin{subfigure}[b]{0.48\textwidth}
        \includegraphics[width=\textwidth]{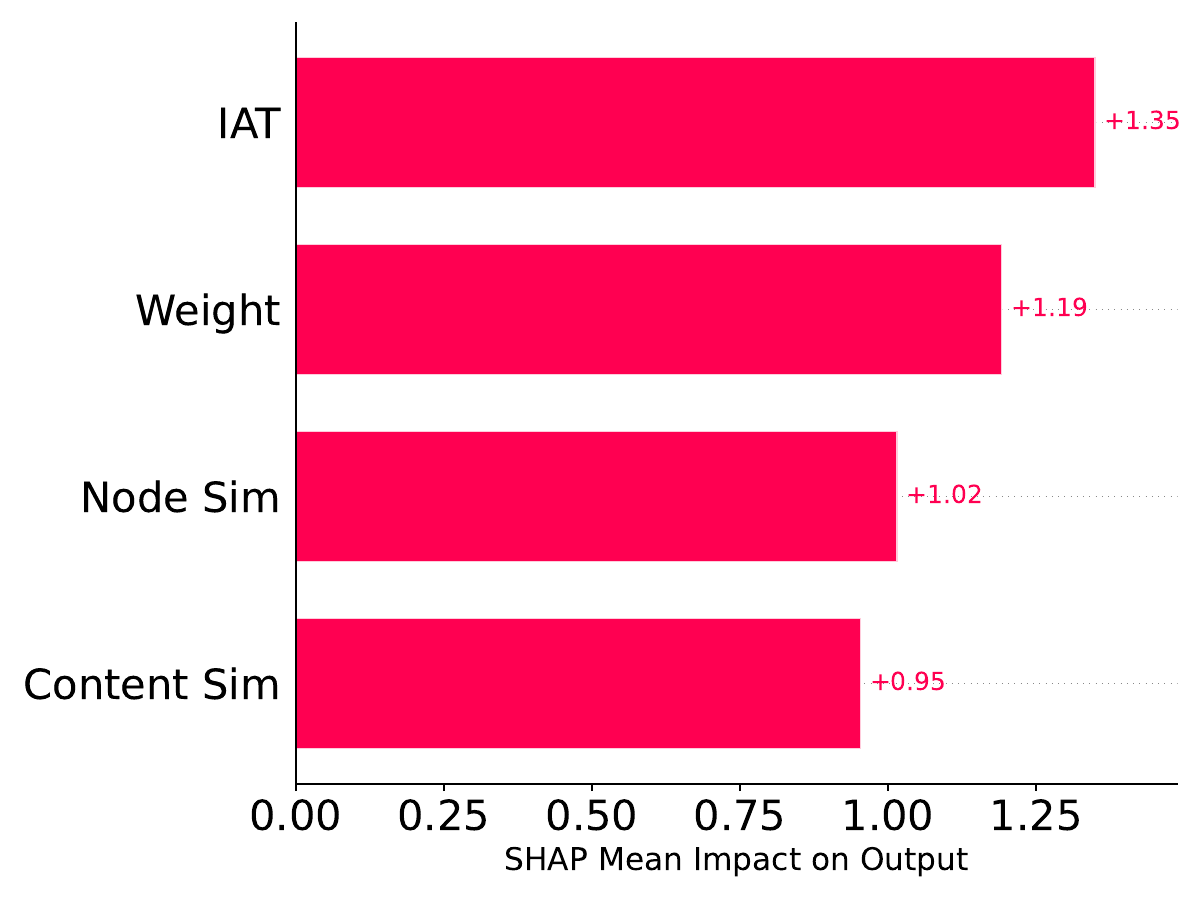}
        \caption{IRA}
        \label{fig:ira_shap}
    \end{subfigure}
    \begin{subfigure}[b]{0.48\textwidth}
        \includegraphics[width=\textwidth]{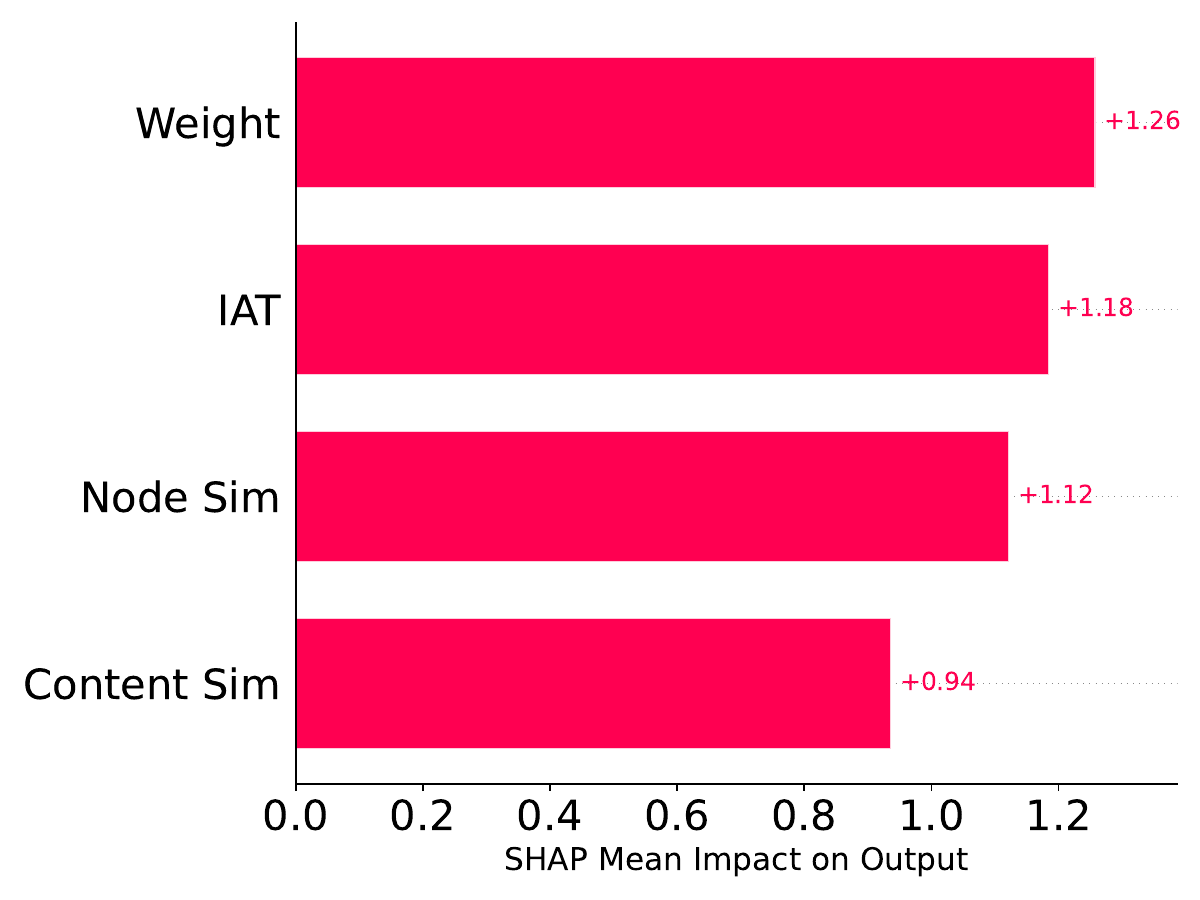}
        \caption{Data Science}
        \label{fig:ds_shap}
    \end{subfigure}
    \caption{Contribution score of each feature variable to the anomalous instances identified by the Isolation Forest model, as estimated by SHAP values. The features are ordered from the highest to the lowest contribution. IAT refers to the inter-arrival time feature. Node Sim refers to the network similarity feature.}
    \label{fig:shap_importance}
\end{figure*}

Isolation forest was applied on the backbone networks to identify anomalous edges/interactions (i.e., those that deviate from the overall distribution of the data).
Specifically, an individual isolation forest model with 100 estimators was trained for each cluster in each dataset. 
The input to the model consisted of four features as described in Section~\ref{sec:features}, which are edge weight, content similarity, inter-arrival time (IAT), and node similarity.
We used the treeSHAP algorithm~\cite{lundberg2018consistent} to compute the SHAP value of each instance within their respective clusters.
The SHAP value measures the contribution of each feature to the overall output of the model, which in this study is the average path length required to reach a data instance. 
Figure~\ref{fig:shap_importance} shows the mean absolute SHAP value of each feature across the population of identified anomalies for each dataset. 
The higher the mean absolute SHAP value of a feature, the more influential the feature is for detecting anomalies. 
We found that the IAT feature had the highest impact for identifying anomalous instances in the IRA while the edge weight feature contributed more to the output of the model in Data Science. 

It is important to note that isolation forest can identify outliers that are present in both tails of the distribution, thus it will classify instances with unusually long inter-arrival times as anomalies.
However, coordinated campaigns typically do not exhibit very large gaps of time between actions (e.g., weeks, months, etc), instead the time difference between actions is shorter but can still be diverse (e.g., seconds, minutes, hours, or even a few days). 
To narrow our focus on the most suspicious instances, we use the median IAT of the distribution of non-anomalous instances as a threshold to remove anomalies with higher IAT values than the norm. 
Table~\ref{tab:anomaly_data} shows the median value of the distribution for each edge feature between anomalous and normal interactions in each cluster and dataset.
We have several observations related to the identified anomalies in each cluster.
First, we found that the clusters of anomalous instances in the IRA exhibit higher content similarity (0.65 $\pm$ 0.15) on average than clusters from DS (0.53 $\pm$ 0.18).
Second, we observed that the IRA clusters exhibit very low IAT values compared to clusters in DS, specifically 4 out of 6 clusters have a median IAT of less than 20 minutes. 
The DS clusters, on the other hand, range from a minimum of 5 hours to a maximum of more than a month. 
This observation is consistent with our expectation that topics discussed in this dataset are less likely to be associated with a specific information campaign. 
Finally, the node similarity feature exhibits relatively higher values in clusters of anomalous interactions in the IRA and two clusters in Data Science.
This suggests that the nodes within these particular clusters are more likely to have similar neighborhoods and co-share hashtags more frequently.

\begin{table*}
    \centering
     \caption{Median value of the distribution of relevant features for anomalous and normal interactions, grouped by previously identified user clusters in each dataset. The inter-arrival time (IAT) is recorded in hours. The content similarity and node similarity is measured using cosine similarity. Anomalous interactions are filtered based on the median IAT of normal interactions.}
    \begin{tabular}{lrrrrrrrrr}
    \toprule
    &\textbf{Cluster}&\multicolumn{2}{c}{\textbf{Weight}}&\multicolumn{2}{c}{\textbf{Content Sim}}&\multicolumn{2}{c}{\textbf{IAT}}&\multicolumn{2}{c}{\textbf{Node Sim}}\\
    \midrule
    &&\textbf{Anom}&\textbf{Norm}&\textbf{Anom}&\textbf{Norm}&\textbf{Anom}&\textbf{Norm}&\textbf{Anom}&\textbf{Norm} \\

    \textbf{IRA}& 1 &21416 & 3561.5 &0.35 &0.33 &0.15 &0.43 &0.45 &0.26 \\
    & 2 &537 & 200&0.82 & 0.82&0.02 & 0.08&0.81 & 0.27\\
    & 3 &114 &10 &0.77 &0.79 &0.02 &0.27 &0.67 &0.25 \\
    & 4 &25 & 5&0.70 & 0.63&122.82 & 1995&0.68 & 0.25 \\
    & 5 &194 & 13&0.59 & 0.46&0.33 &147.4 &0.62 &0.26 \\
    & 6 &408 & 69&0.66 & 0.49&3.16 &107.68 &0.42 &0.25 \\
    \midrule
    \textbf{DS}& 1 &33 &21 &0.46 &0.49 &440.04 &1053.38 &0.53 &0.30 \\
    & 2 &2305 &288 &0.47 &0.44 &5.47 &69.69 &0.82 &0.33 \\
    & 3 &135 &69 &0.47 &0.46 &100.08 &218.48 &0.63 &0.30 \\
    & 4 &9 &2 &0.46 &0.35 &986.05 &4988.58 &0.30 &0.26 \\
    & 5 &8 &3 &0.46 &0.38 &1420.33 &5528.77 &0.37 &0.32 \\
    \bottomrule
    
    \end{tabular}
    \label{tab:anomaly_data}
\end{table*}

\subsection{Qualitative Analysis of Anomalous Clusters}

We examined the content shared by anomalous users within each cluster, as identified by our framework.
We use topic modeling to identify the general themes and topics of discussion within each cluster, as well as analyze the most frequently shared hashtags by these users.
Our goal is to identify which clusters are likely related to suspicious activity and potential coordination.

In the DS dataset, all clusters exhibited similar behavior and characteristics.
The discussions revolve around advancements in data science and its growing importance in various fields. 
Tweets primarily share information about resources, opportunities, applications and the overall impact of data science.
We observed that the inter-arrival times (ranging from a minimum of 5 hours to a maximum of 59 days) and content similarity (ranging from a minimum of 0.46 to a maximum of 0.47) across clusters likely indicate no evidence of coordinated behavior.
The discussions appear to be organic and align with the behavior we expected for this dataset.

In the IRA dataset, we grouped the identified user clusters into three categories: \textit{News Feed}, \textit{Pro-Russian Nationalistic Voices}, and \textit{Fear-mongers and Trolls}. 
Our observations for this dataset align with some of the categories previously identified in the work of Linvil et al.~\cite{linvill2020troll}, which unlike our framework, heavily relies on several qualitative analyses for cluster identification.

\textbf{News Feed (Cluster 1 and 4)}
Cluster 1 consists of 13 accounts and a total of 21 edges, indicating a relatively small network.
The cluster exhibits a very high level of co-hashtag promotion as evidenced by the edge weight of co-shares (with a median of 21,416).
There is a moderate degree of similarity in the users' posts with a median cosine similarity of 0.35.
The median inter-arrival time between actions is relatively short (around 9 minutes).
The main topics of discussion were around news related to political events on a global scale.
The analysis of the most frequently shared hashtags revealed the presence of hashtags such as \#news, \#local, \#politics, \#sports, and \#entertainment, which indicates a broad coverage of topics.
Cluster 4, consisting of 703 users and 8,892 edges, also engaged in actively sharing news content, but especially related to Russia.
The tweets in this cluster exhibit a broad coverage of news from different regions within Russia as seen by the presence of hashtags referring to specific Russian cities such as \#UFA, \#SPB, \#Yaroslavl, and \#Voronezh.
Contrary to cluster 1, cluster 4 does not exhibit strong indications of synchronized behavior. 
Its inter-arrival time between actions is long, with a median of around 5 days, which suggest more sporadic engagement compared to other clusters.

\textbf{Pro-Russian Nationalistic Voices (Clusters 2 and 3)}
Cluster 2 consists of 422 users and 3,213 edges while cluster 3 consists of 346 users and 1,283 edges.
Both clusters exhibit high levels of content similarity, with cluster 2 having a median cosine similarity of 0.82 and cluster 3 with 0.77.
The inter-arrival time between actions is relatively short for both clusters, with 1 minute and 1.5 minutes for cluster 2 and 3, respectively. 
There is also high similarity in connections among users as seen by a node similarity of 0.81 for cluster 2 and 0.67 for cluster 3.
The content shared within both clusters primarily focuses on the ongoing conflict between Ukraine and Russia, with messages criticizing foreign policies, particularly those of Western countries. 
Cluster 2 engages in discussions related to the downing of the Malaysian airlines flight in 2014 and makes claims regarding Ukraine's involvement. 
Both clusters actively promote a sense of national identity in Russia through their tweets.
Cluster 2 frequently promotes hashtags such as \#RussianSpirit, \#KievShotDowntheBoeing, \#KievTellTheTruth, while cluster 3 promotes hashtags such as \#AmericanPlague, \#AgainstSanctions, and \#MadeInRussia.
The high levels of content similarity, short inter-arrival times, and presence of hashtags with provocative allegations suggest a likely coordinated effort to disseminate specific narratives focused on the Ukraine-Russia conflict. 

\textbf{Fear-mongers and Trolls (Cluster 5 and 6)}
Cluster 5 consists of 735 users and 6,220 edges and cluster 6 consists of 710 users and 9,234 edges.
These clusters present behaviors and characteristics that strongly indicate suspicious and coordinated behavior.
In cluster 5, we observed that messages are mostly around the 2016 US presidential elections, with users frequently sharing hashtags such as \#WakeUpAmerica, \#tcot (Top Conservatives on Twitter), and \#pjnet (Patriot Journalist Network).
The cluster also targets cultural identity and social issues as seen by the sharing of hashtags such as \#BlackLivesMatter, \#IslamKills, and \#IslamistIsTheProblem.
The median content similarity is 0.59, which is a moderate level of similarity, and the median inter-arrival time was 20 minutes. 
Cluster 6, on the other hand, engages in pushing fabricated crisis events.
Some were related to nuclear incidents as indicated by the frequent sharing of hashtags such as \#Fukushima2015 and \#ColumbianChecmicals.
Another frequently pushed story was related to Koch Farms during the Thanksgiving of 2015.
The story alleged that Koch Farms' turkey production was contaminated with salmonella, resulting in severe food poisoning.
The content similarity in cluster 6 was 0.66 and with slightly higher inter-arrival times of approximately 3 hours.
Our observations for cluster 5  and 6 align with the \textit{Trolls} and \textit{Fearmonger} categories identified in~\cite{linvill2020troll}. 
Overall, the content shared in these clusters, which are politically divisive and contain inflammatory messages, along with their particular characteristics suggest a strong presence of coordinated information campaigns.

\section{Conclusions and Future Work}
\label{sec:conclusions}

This study proposes a multifaceted framework to detect user accounts possibly involved in coordinated activity on social media platforms.
Our approach involves the analysis of content, timing, and network dimensions to distinguish between organic behavior and suspicious coordinated operations.
Our analyses revealed the following observations.

First, we demonstrated the effectiveness of our framework in isolating organic behavior from inorganic behavior across discussions on two Twitter datasets of known coordinated information operations to likely only organic discussions.
Second, we showed that our backbone extraction component proved valuable in reducing the search space within the original co-sharing networks, revealing the underlying core structures.
Third, our anomaly detection model based on isolation forests effectively identified anomalous instances, and combined with explainable methods like SHAP, it provides additional insights into the contribution and importance of each feature.
Finally, we identified and characterized clusters of users who likely engaged in coordinated campaigns based on their shared content and time locality. 

However, our framework does have some limitations some by choice others due to the nature of social media data.
First, we rely on certain assumptions such as the prevalence of organic behavior outweighing inorganic behavior.
As campaigns evolve and become more sophisticated (e.g., mimicking human behavior better), it may become increasingly challenging to detect anomalies.  
Second, the absence of ground truth data on coordinated behaviors for some of our datasets makes it difficult to claim with certainty the presence of an actual coordinated effort.
Our claims are based solely on the observed characteristics and patterns within the data, thus further research is necessary to confirm the extent to which particular accounts are linked to the coordinated campaign. 
Third, data accessibility is a pressing concern for studies aiming to identify nefarious processes on social media platforms. 
Recently, some platforms have restricted their APIs or have shifted towards paid models for data access, making it challenging for researchers to obtain valuable data.
Fourth, our analysis does not consider reposting behavior (e.g., retweets) or user engagement (e.g., replies) as we only focus on coordinated information promotion.
Incorporating this information in our framework can provide insights on the actual scale of the campaign, its reach, and its impact on online discussions.
Fifth, while our framework can effectively isolate unusual interactions, it cannot automatically detect if clusters are part of a coordinated inauthentic campaign.
Qualitative analysis is still needed to determine the level of inauthenticity in these interactions. 
In future work, we could create random baselines that disrupt temporal relationships between users.
This would help in assessing how much the multivariate distributions of observed coordinated features deviate from chance.

Future work will also focus on augmenting the features used for anomaly detection to make them more adaptable with evolving coordination strategies.
Another venue for future work is the detection of coordinated campaigns in multi-platform settings, where accounts on multiple social media platforms promote a shared agenda.

%
%
%
\bibliographystyle{splncs04}
\bibliography{bibtex}
\end{document}